\documentclass[twocolumn,superscriptaddress,preprintnumbers,amsmath,amssymb,prl]{revtex4}

\usepackage[dvipdfmx]{graphicx}
\usepackage{dcolumn}
\usepackage{color}
\usepackage{bm}
\usepackage{booktabs}
\usepackage{braket}

\newcommand{\upa}{\uparrow}
\newcommand{\dna}{\downarrow}

\newcommand{\ulin}{\underline}

\newcommand{\beeq}{\begin{eqnarray}}
\newcommand{\eneq}{\end{eqnarray}}

\begin{document}


\title{Emergence of multiple Higgs modes due to spontaneous breakdown of a $\mathbb{Z}_2$ symmetry in a superconductor}

\author{Shunji Tsuchiya}
\email{tsuchiya@phys.chuo-u.ac.jp}
\affiliation{Department of Physics, Chuo University, 1-13-27 Kasuga, Bunkyo-ku, Tokyo 112-8551, Japan}

\date{\today}

\begin{abstract}
We study the Higgs mode in a Bardeen-Cooper-Schrieffer (BCS) superconductor. Motivated by the observation that U(1) symmetry of the BCS Hamiltonian is not essential for the Higgs mode, we study the Ising-like Hamiltonian in the pseudospin representation.
We show that the Higgs mode emerges as the lowest excited state of the Ising-like Hamiltonian due to spontaneous breakdown of $\mathbb{Z}_2$ symmetry under the time-reversal operation $\mathcal T$ in the pseudospin space.
We further predict the existence of multiple Higgs modes that have quantized energy $2(n+1)\Delta_0$ ($0\le n\le N_{k_F}$), where $\Delta_0$ is the superconducting gap, $n$ is an integer, and $N_{k_F}$ is the number of states on the Fermi surface. 

\end{abstract}
\maketitle

{\it Introduction.}--
Higgs modes and Nambu-Goldstone (NG) modes are ubiquitous collective excitations that arise in systems with spontaneous symmetry breaking \cite{anderson-58,nambu-60,goldstone-61,goldstone-62,higgs-64}. They are amplitude and phase modes of the order parameter and appear in a broad class of systems in condensed-matter and particle physics. 
Stimulated by the discovery of the Higgs boson in the Standard Model \cite{cms-12}, interests in Higgs modes in condensed-matter physics have grown recently \cite{pekker-15,shimano-20}. Higgs modes have been observed in various condensed-matter systems including superconductors \cite{sooryakumar-81,matsunaga-13,matsunaga-14,sherman-15,katsumi-18,krull-16,chu-20}, superfluids \cite{giannetta-80,zavjalov-16,nguyen-19}, quantum spin systems \cite{ruegg-08,jain-17,hong-17,souliou-17}, charge-density-wave (CDW) materials \cite{demsar-99,yusupov-10}, and ultracold atomic gases \cite{bissbort-11,endres-12,leonard-17,behrle-18}.
\par
Although Higgs modes are believed to emerge with NG modes when continuous symmetries are spontaneously broken, fundamental questions on Higgs modes remain to be understood. Namely, in contrast with NG modes, whose existence is predicted by the Goldstone theorem \cite{goldstone-62}, Higgs modes do not necessarily appear in systems exhibiting spontaneous breakdown of continuous symmetries. 
For instance, whereas a Higgs mode appears in a Bardeen-Cooper-Schrieffer (BCS) superconductor \cite{littlewood-81}, it disappears in a Bose-Einstein condensate (BEC) \cite{varma-02}, although spontaneous breakdown of U(1) symmetry occurs in both systems and, furthermore, the former continuously evolves to the latter in the BCS-BEC crossover phenomenon \cite{leggett-80,nozieres-85,ohashi-02,regal-04}.
\par
Particle-hole (p-h) symmetry is a crucial condition for the emergence of Higgs modes.
In superconductors, for instance, Higgs modes appear when the fermion energy dispersion is p-h symmetric \cite{engelbrecht-97,varma-02,tsuchiya-13,cea-15}. 
In a bosonic superfluid in an optical lattice, the system exhibits the approximate p-h symmetry in the vicinity of the tip of the Mott lobe, where a Higgs mode appears \cite{sachdev-98,endres-12,nakayama-15,diliberto-18,nakayama-19}. 
\par
In the previous work \cite{tsuchiya-18}, we have shown that the BCS Hamiltonian for a superconductor with a p-h symmetric fermion energy dispersion has the non-trivial $\mathbb{Z}_2$ symmetries under discrete operations. We refer to them as the charge-conjugation ($\mathcal C$), parity ($\mathcal P$), and time-reversal ($\mathcal T$) operations in analogy with the corresponding ones in the relativistic field theory \cite{lee-81}.
We found that $\mathcal T$ and $\mathcal P$ are spontaneously broken, while $\mathcal C$ is unbroken in the BCS ground state. We further conjectured that the emergence of the Higgs mode may be a consequence of the spontaneous breakdown of $\mathbb{Z}_2$ symmetry under $\mathcal T$.  
\par
\begin{figure}
\centering
\includegraphics[width=\linewidth]{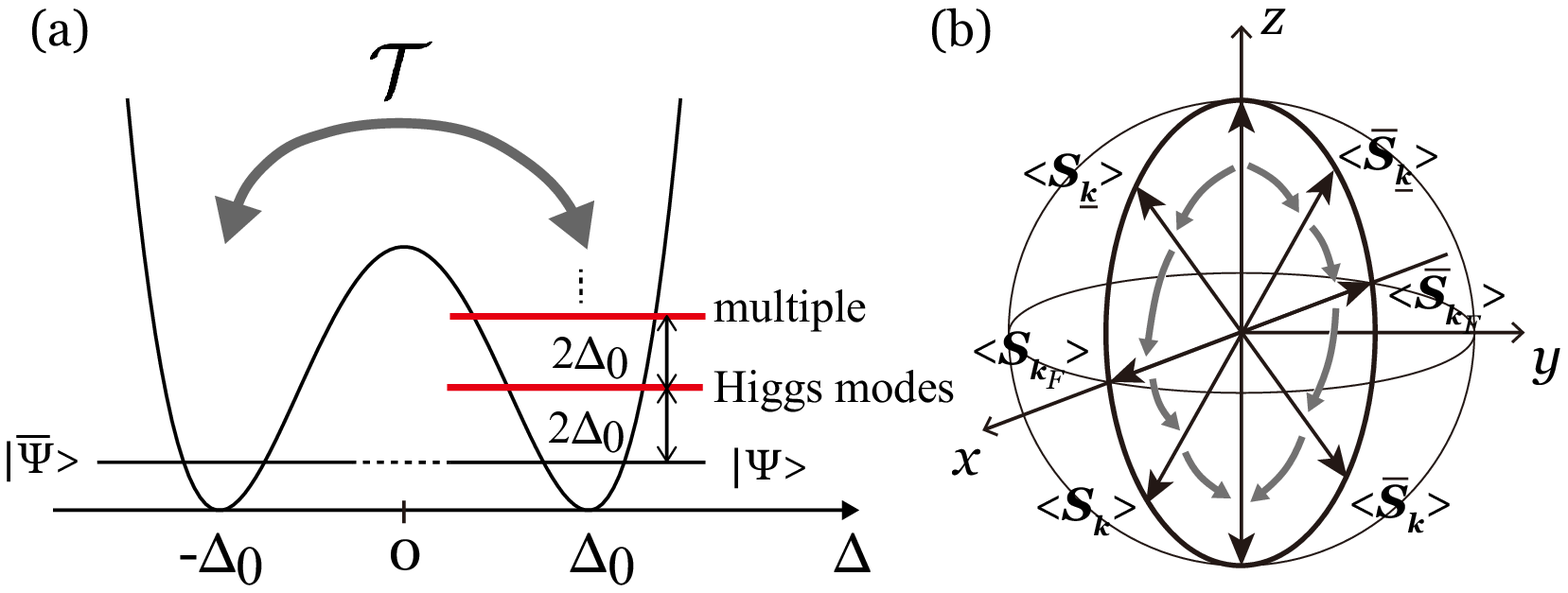}
\caption{(a) Schematic illustration of the effective double-well potential for $\mathcal H_I$. $\ket{\Psi}$ ($\ket{\bar \Psi}$) is the SC ground state for the gap function $\Delta_0$ ($-\Delta_0$). The multiple Higgs modes with excitation energy $2(n+1)\Delta_0$ ($0\leq n\leq N_{k_F}$) emerge due to spontaneous breakdown of $\mathbb{Z}_2$ symmetry under $\mathcal T$ in the ground state $\ket{\Psi}$. (b) Pseudospin configuration for the SC ground state with broken $\mathbb{Z}_2$ symmetry under $\mathcal T$. The pseudospins for $\ket{\Psi}$ ($\ket{\bar \Psi}$) turn towards positive (negative) $x$-direction as $\bm k$ evolves from below to above the FS. We denote $\langle\bm S_{\bm k}\rangle=\bra{\Psi}{\bm S}_{\bm k}\ket{\Psi}$ and $\langle\bm \bar{\bm S}_{\bm k}\rangle=\bra{\bar\Psi}{\bm S}_{\bm k}\ket{\bar\Psi}$.
}
\label{fig.doublewell_spin}
\end{figure}

In this Letter, extending the previous work, we establish that the Higgs mode emerges in a superconductor due to {\it spontaneous breakdown of $\mathbb{Z}_2$ symmetry under $\mathcal T$}.
Motivated by the observation that U(1) symmetry is not essential for the Higgs mode, we study the Ising-like Hamiltonian in the pseudospin representation $\mathcal H_I$ derived from the BCS Hamiltonian, which exhibits $\mathbb{Z}_2$ symmetries under $\mathcal C$, $\mathcal P$, and $\mathcal T$.
We show that the Higgs mode appears as the lowest excited state of $\mathcal H_I$ due to spontaneous breakdown of $\mathbb{Z}_2$ symmetry under $\mathcal T$ in the ground state. 
Furthermore, we predict the existence of the multiple Higgs modes that have quantized energy $2(n+1)\Delta_0$ ($0\le n\le N_{k_F}$) and include the original one ($n=0$), where $\Delta_0$ is the superconducting (SC) gap, $n$ is an integer, and $N_{k_F}$ is the number of states on the Fermi surface (FS).
Our main results are schematically illustrated in Fig.~\ref{fig.doublewell_spin} (a).
\par
{\it Particle-hole symmetry and Higgs modes.}--To illustrate the condition in which the Higgs mode is well-defined in a superconductor, let us first discuss the relation between p-h symmetry and the Higgs mode. For simplicity, we consider the BCS Hamiltonian in the pseudospin representation \cite{anderson-58,tsuchiya-18}
\begin{eqnarray}
  \mathcal H_{\rm BCS}&=&\mathcal H_K+\mathcal H_{XY}~,\quad
  \mathcal H_K=\sum_{\bm k}2\xi_{\bm k} S_{z\bm k}~,\\
  \mathcal H_{XY}&=&-g\sum_{\bm k,\bm k'}(S_{x\bm k}S_{x\bm k'}+S_{y\bm k}S_{y\bm k'})~,
\label{eq.HBCS}
\end{eqnarray}
where $\bm S_{\bm k}=(S_{x\bm k}, S_{y\bm k}, S_{z\bm k})$ is the spin-1/2 pseudospin operator and $S_\mu=\sum_{\bm k}S_{\mu\bm k}$ is the total spin. We set $\hbar=1$ throughout the Letter.
The fermion vacuum corresponds to the spin-down state ($\ket{0}_{\bm k}=\ket{\downarrow}_{\bm k}$) and the fully occupied state to the spin-up state ($c_{\bm k\uparrow}^\dagger c_{-\bm k\downarrow}^\dagger\ket{0}_{\bm k}=\ket{\uparrow}_{\bm k}$), where $c_{\bm ks}^\dagger$ is the creation operator for a fermion with momentum $\bm k$ and spin $s$($=\upa,\dna$). $\mathcal H_K$ is the kinetic energy term and $\mathcal H_{XY}$ is the interaction term. Here, $\xi_{\bm k}=\varepsilon_{\bm k}-\mu$ is the energy dispersion of a fermion measured from the chemical potential $\mu$ and $g(>0)$ is the coupling constant for the attractive interaction between fermions. We do not specify the form of $\varepsilon_{\bm k}$ for generality of argument.  
\par
$\mathcal H_{\rm BCS}$ has rotational U(1) symmetry about the $z$-axis in the pseudospin space that reflects U(1) gauge symmetry. It is spontaneously broken in the SC state \cite{anderson-58,nambu-60}. 
When $\xi_{\bm k}$ satisfies the condition for the p-h symmetry
\begin{equation}
-\xi_{\ulin{\bm k}}=\xi_{\bm k}~,\label{eq.phcondition}
\end{equation}
$\mathcal H_{\rm BCS}$ has additional $\mathbb{Z}_2$ symmetries under $\mathcal C$ and $\mathcal T$ \cite{tsuchiya-18}. 
Here, $\bm k$ and $\underline{\bm k}$ are a pair of wave vectors that are located on the opposite side of the FS with the same distance from it (See Figs.~1~(a)-(c) in Ref.~\cite{tsuchiya-18}). $\mathcal C$, $\mathcal P$, and $\mathcal T$ are defined as
\begin{eqnarray}
  &&\mathcal C={\mathcal F}\prod_{\bm k}\sigma_{x\bm k}~,\quad \mathcal F=\sideset{}{^{'}}\prod_{\bm k} f_{\bm k,\underline{\bm k}}~,\label{eq.operatorC}\\
  &&\mathcal T=U_T\mathcal K~,\quad U_T=\mathcal F\prod_{\bm k}(-i\sigma_{y\bm k})\label{eq.operatorT}~,\\
  &&\mathcal P=\prod_{\bm k}\sigma_{z\bm k}~,
\end{eqnarray}
where $\prod_{\bm k}'\equiv\prod_{\xi_{\bm k}>0}$, $\sigma_{\mu\bm k}\equiv 2S_{\mu\bm k}$, $f_{\bm k\ulin{\bm k}}$ is a swapping operator between the states of $\bm k$ and $\ulin{\bm k}$, and $\mathcal K$ is the complex conjugation operator. Note that $\mathcal P$ represents a $\pi$ rotation about the $z$-axis that is an element of U(1).
$\mathcal{CPT}$ and all other permutations of $\mathcal C$, $\mathcal P$, and $\mathcal T$ are exact symmetries.
$\mathbb Z_2$ symmetries under $\mathcal T$ and $\mathcal P$ are spontaneously broken with U(1) in the SC state, while that under $\mathcal C$ is unbroken \cite{tsuchiya-18}.
\par
On the other hand, the Higgs mode appears as an amplitude mode only when $\xi_{\bm k}$ is p-h symmetric, i.e., it satisfies Eq.~(\ref{eq.phcondition}) \cite{varma-02,tsuchiya-18}.
In the classical spin analysis, for a gap function in the ground state $\Delta_0>0$, amplitude and phase fluctuations of the gap function are proportional to those of the $x$ and $y$ components of the total spin as $\delta\Delta\propto\delta S_x$ and $\delta\theta\propto\delta S_y$, respectively. 
$\delta \Delta$ and $\delta \theta$ are uncoupled and pure amplitude oscillations of the Higgs mode are allowed only when $\xi_{\bm k}$ is p-h symmetric, because the off-diagonal element of the dynamical spin susceptibility $\chi_{xy}(\omega)$ vanishes due to the opposite parity of $S_x$ and $S_y$ under $\mathcal C$ \cite{tsuchiya-18,supplement}. 
When $\xi_{\bm k}$ is not p-h symmetric, finite $\chi_{xy}(\omega)$ couples amplitude and phase fluctuations.
\par
Moreover, the Higgs mode is stable only when $\xi_{\bm k}$ is p-h symmetric \cite{tsuchiya-18}. It is prohibited to decay into single-particle excitations despite its degeneracy with the lower edge of the single-particle continuum at $2\Delta_0$, because the Higgs mode has even parity and the single-particle excitations at $2\Delta_0$ have odd parity under $\mathcal C$. 
When $\xi_{\bm k}$ is not p-h symmetric, however, the Higgs mode merges with the single-particle continuum and suffers from strong damping due to decay into single-particle excitations.
Thus, the Higgs mode is well-defined only when $\mathbb Z_2$ under $\mathcal T$ is spontaneously broken with U(1) in the SC state. 
\par
{\it Minimal Hamiltonian for Higgs modes.}--
To argue the origin of the Higgs mode, we focus on a p-h symmetric system, in which the Higgs mode is well-defined, and assume Eq.~(\ref{eq.phcondition}) in the rest of the Letter.
\par 
In the classical spin analysis, one finds that the $y$ component of the interaction term $\mathcal H_Y=-gS_y^2$ may be irrelevant to the Higgs mode, because the Higgs mode involves out-of-phase oscillations of $\delta S_{y\bm k}$ and $\delta S_{y\ulin{\bm k}}$, whereas $\mathcal H_Y$ induces in-phase oscillations of them~\cite{anderson-58,tsuchiya-18, supplement}. 
Furthermore, the analysis based on the Holstein-Primakoff (H-P) theory suggests that $\mathcal H_Y$ is not necessary for constructing the creation operator of the Higgs mode~\cite{tsuchiya-18}.
Motivated by these observations, we neglect $\mathcal H_Y$ and study the Ising-like Hamiltonian $\mathcal H_I$:
\begin{eqnarray}
&&\mathcal H_I=\mathcal H_{K}+\mathcal H_{X}~,\label{eq.HI}\\
&&\mathcal H_{X}=-g\sum_{\bm k,\bm k'} S_{x\bm k}S_{x\bm k'}=-gS_x^2~.
\label{eq.HX}
\end{eqnarray}
Note that $\mathcal H_I$ is invariant under $\mathcal C$, $\mathcal P$, and $\mathcal T$, despite it loses U(1) symmetry. We derive the Higgs mode from $\mathcal H_I$ to demonstrate that the emergence of the Higgs mode is not due to spontaneous breakdown of U(1) but to that of $\mathbb Z_2$ under $\mathcal T$.
\par
{\it Discrete symmetries of $\mathcal H_I$.}--
In addition to $\mathbb Z_2$ symmetries under $\mathcal C$, $\mathcal P$, and $\mathcal T$, $\mathcal H_I$ has $\mathbb Z_2$ symmetries under local charge-conjugation operations in momentum space ${\mathcal C}_{\bm k\ulin{\bm k}}=f_{\bm k,\underline{\bm k}}\sigma_{x\bm k}\sigma_{x\underline{\bm k}}$ and ${\mathcal C}_{\bm k_F}=\sigma_{x\bm k_F}$, where $\mathcal C$ can be written as $\mathcal C=\prod_{\bm k}'\mathcal C_{\bm k\ulin{\bm k}}\otimes\prod_{\bm k_F}{\mathcal C}_{\bm k_F}$.
To prove the invariance under ${\mathcal C}_{\bm k\ulin{\bm k}}$, it is convenient to introduce a pseudospin operator
\begin{eqnarray}
  S_{\mu\bm k\underline{\bm k}}=S_{\mu\bm k}-(-1)^{\delta_{\mu,x}}S_{\mu\ulin{\bm k}}~,~(\mu=x,y,z),\label{eq.Skkbar}
\end{eqnarray} 
where $\bm k$ is above the FS ($\xi_{\bm k}>0$). It commutes with $\mathcal C_{\bm k\ulin{\bm k}}$.
\par
The kinetic energy term can be written as
\begin{eqnarray}
\mathcal H_{K}=\sideset{}{^{'}}\sum_{\bm k}2\xi_{\bm k}S_{z\bm k\underline{\bm k}}~,
\label{eq.H_K1}
\end{eqnarray}
where $\sum_{\bm k}'\equiv\sum_{\xi_{\bm k}>0}$. We obtain $[\mathcal H_K,{\mathcal C}_{\bm k\ulin{\bm k}}]=[\mathcal H_K,{\mathcal C}_{\bm k_F}]=0$ from Eq.~(\ref{eq.H_K1}). Since $S_x=\sum_{\bm k}'S_{x\bm k\ulin{\bm k}}+\sum_{\bm k_F}S_{x\bm k_F}$ commutes with either $\mathcal C_{\bm k\ulin{\bm k}}$ and $\mathcal C_{\bm k_F}$, we also obtain $[\mathcal H_X,{\mathcal C}_{\bm k\ulin{\bm k}}]=[\mathcal H_X,{\mathcal C}_{\bm k_F}]=0$ and $[\mathcal H_I,{\mathcal C}_{\bm k\ulin{\bm k}}]=[\mathcal H_I,{\mathcal C}_{\bm k_F}]=0$.
Consequently, $\mathcal H_I$ has $\mathbb Z_2$ symmetry under each of ${\mathcal C}_{\bm k\ulin{\bm k}}$ and ${\mathcal C}_{\bm k_F}$.
$\mathcal H_Y$ and $\mathcal H_{\rm BCS}$, in contrast, commute with neither ${\mathcal C}_{\bm k\ulin{\bm k}}$ nor ${\mathcal C}_{\bm k_F}$.
\par
{\it Discrete symmetries in low-energy states of $\mathcal H_I$.}--
We next study the symmetry of the ground state of $\mathcal H_I$. 
The pseudospin configuration of it is shown in Fig.~\ref{fig.doublewell_spin} (b): The pseudospins turn smoothly from up to down towards either  positive or negative $x$-direction as $\bm k$ evolves from below to above the FS, where the gap function $\Delta\equiv g\langle S_x\rangle$ is positive and negative in the former and latter cases, respectively.
One of them being chosen spontaneously, $\mathbb Z_2$ under $\mathcal P$ is broken in the ground state.
It leads to breakdown of $\mathbb Z_2$ under $\mathcal T$ due to unbroken $\mathbb Z_2$ under $\mathcal C$, which is shown below, and the exact symmetry under $\mathcal C\mathcal P\mathcal T$. 
The energy landscape of $\mathcal H_I$ is effectively described by the double-well potential, as schematically illustrated in Fig.~\ref{fig.doublewell_spin} (a).
\par
Let us show that $\mathbb Z_2$ under each of $\mathcal C_{\bm k\ulin{\bm k}}$, $\mathcal C_{\bm k_F}$, and $\mathcal C$ is unbroken in the ground state. 
If we introduce a gap function $\Delta=\Delta_0>0$, $\mathcal H_I$ reduces to the mean-field (MF) Hamiltonian
\begin{eqnarray}
\mathcal H_{\rm MF}=\sum_{\bm k}2E_{\bm k}S'_{z\bm k}=\sideset{}{^{'}}\sum_{\bm k}2E_{\bm k}S'_{z\bm k\ulin{\bm k}}-2\Delta_0\sum_{\bm k_F}S_{x\bm k_F},\label{eq.HMF1}
\end{eqnarray}
where $E_{\bm k}=\sqrt{\xi_{\bm k}^2+\Delta_0^2}$ is the dispersion of a single-particle excitation referred to as a bogolon. 
$\bm S'_{\bm k}$ is the pseudospin operator for a pair of bogolons with $\ket{\upa'}_{\bm k}$ and $\ket{\dna'}_{\bm k}$ being the eigenstates of $S'_{z\bm k}$ \cite{anderson-58,tsuchiya-18,supplement}.
Note that $\mathcal H_I$ and $\mathcal H_{\rm BCS}$ reduce to the same MF Hamiltonian for a real gap function.  
Here, we have introduced a spin operator
\begin{eqnarray}
S_{\mu\bm k\ulin{\bm k}}'=S_{\mu\bm k}'-(-1)^{\delta_{z,\mu}}S_{\mu\ulin{\bm k}}'~,~(\mu=x,y,z),\label{eq.Sd}
\end{eqnarray}
where $\bm k$ is above the FS ($\xi_{\bm k}>0$). 
It commutes with $\mathcal C_{\bm k\ulin{\bm k}}$. Using Eq.~(\ref{eq.HMF1}), we obtain $[\mathcal H_{\rm MF},\mathcal C_{\bm k\ulin{\bm k}}]=[\mathcal H_{\rm MF},\mathcal C_{\bm k_F}]=0$. Consequently, $\mathbb{Z}_2$ symmetry under each of $\mathcal C_{\bm k\ulin{\bm k}}$ and $\mathcal C_{\bm k_F}$ as well as under $\mathcal C$ is unbroken in the ground state of $\mathcal H_I$.
\par
The eigenstates of $\mathcal H_{\rm MF}$ can be characterized by parity under $\mathcal C_{\bm k\ulin{\bm k}}$ and $\mathcal C_{\bm k_F}$.
The ground state of Eq.~(\ref{eq.HMF1}) $\ket{\Psi}=\prod_{\bm k}'\ket{-1'}_{\bm k\ulin{\bm k}}\otimes\prod_{\bm k_F}\ket{+}_{\bm k_F}$ has even parity for all $\mathcal C_{\bm k\ulin{\bm k}}$ and $\mathcal C_{\bm k_F}$, where $\ket{\pm}_{\bm k}\equiv(\ket{\uparrow}_{\bm k}\pm\ket{\downarrow}_{\bm k})/\sqrt{2}$. Here, the product states $\ket{s',t'}_{\bm k\ulin{\bm k}}=\ket{s'}_{\bm k}\otimes\ket{t'}_{\ulin{\bm k}}$ 
($s,t=\uparrow,\downarrow$) are decomposed into the even and odd parity states under $\mathcal C_{\bm k\ulin{\bm k}}$, where the former consist of the spin-1 triplet states $\ket{m'}_{\bm k\ulin{\bm k}}$ ($m=1,0,-1$) for $S'_{z\bm k\underline{\bm k}}$  and the latter the spin-0 singlet state $\ket{\tilde{0}'}_{\bm k\ulin{\bm k}}$ \cite{supplement}.
The ground state $\ket{\bar\Psi}=\mathcal T\ket{\Psi}$ for $\Delta=-\Delta_0$ has even and odd parity under $\mathcal C_{\bm k\ulin{\bm k}}$ and $\mathcal C_{\bm k_F}$, respectively.
\par
Parity of single-particle states plays a key role for the stability of the Higgs mode discussed below. $\ket{e_{\bm k\ulin{\bm k}}^l}=\ket{l'}_{\bm k\ulin{\bm k}}\otimes\prod_{\bm k'\neq\bm k}'\ket{-1'}_{\bm k'\ulin{\bm k}'}\otimes\prod_{\bm k_F}\ket{+}_{\bm k_F}$ ($l=0,\tilde{0}$) are degenerate single-particle states with excitation energy $2E_{\bm k}$ that form the two-particle continuum. They have even ($l=0$) and odd ($l=\tilde 0$) parity under $\mathcal C_{\bm k\ulin{\bm k}}$. They both have even parity under $\mathcal C_{\bm k_F}$.
$\ket{e_{\bm k_F}}=\ket{-}_{\bm k_F}\otimes\prod_{\bm k}'\ket{-1'}_{\bm k\ulin{\bm k}}\otimes\prod_{\bm k_F'\neq\bm k_F}\ket{+}_{\bm k_F'}$ forms the lower edge of the two-particle continuum at $2\Delta_0$ that is degenerate with the Higgs mode. It has odd parity under $\mathcal C_{\bm k_F}$ and $\mathcal C$. 
\par
{\it Emergence of Higgs mode due to breakdown of $\mathbb Z_2$ under $\mathcal T$}.-- 
To derive the Higgs mode, we apply the H-P theory to $\mathcal H_I$ \cite{holstein-40,supplement}. 
The second-order term in spin fluctuation can be diagonalized as $\mathcal H_I^{(2)}=\omega_{\rm H}\beta_{\rm H}^\dagger\beta_{\rm H}$. The collective mode has excitation energy $\omega_{\rm H}=2\Delta_0$ and its creation operator is given as
\begin{eqnarray}
\beta_{\rm H}^\dagger=\sideset{}{^{'}}\sum_{\bm k}\frac{\xi_{\bm k}}{E_{\bm k}}\left(\frac{S_{\bm k\ulin{\bm k}}^{'+}}{2|\Delta_0|-2E_{\bm k}}+\frac{S_{\bm k\ulin{\bm k}}^{'-}}{2|\Delta_0|+2E_{\bm k}}\right)~,\label{eq.betaH}
\end{eqnarray}
where $S_{\bm k\ulin{\bm k}}^{'\pm}=S'_{x\bm k\ulin{\bm k}}\pm iS'_{y\bm k\ulin{\bm k}}$.
We find that $\beta_{\rm H}^\dagger$ coincides exactly with the creation operator of the Higgs mode derived from $\mathcal H_{\rm BCS}$ (See Eq.~(27) in Ref.~\cite{tsuchiya-18}). Thus, $\beta_{\rm H}^\dagger$ represents the Higgs mode. The NG mode does not appear, because $\mathcal H_I$ does not have U(1) symmetry.  
\par
The emergence of the Higgs mode from $\mathcal H_I$ clearly shows that U(1) symmetry breaking is not essential for it. Now, the origin of the Higgs mode can be attributed to the spontaneous breakdown of $\mathcal T$ or $\mathcal P$.
However, the Higgs mode disappears, for example, in the BEC regime of the BCS-BEC crossover, despite $\mathcal P$ is broken with U(1) in the SC phase \cite{engelbrecht-97}.
Therefore, the emergence of the Higgs mode is considered due to breakdown of ${\mathbb Z}_2$ symmetry under $\mathcal T$.
This conclusion, together with the fact that the Higgs mode arises from $\mathcal H_{\rm BCS}$, indicates that the SC phase transition for $\mathcal H_{\rm BCS}$ with a p-h symmetric $\xi_{\bm k}$ is associated with breakdown of not U(1), but ${\mathbb Z}_2$ under $\mathcal T$. Namely, $\mathcal H_{\rm BCS}$ effectively reduces to $\mathcal H_I$ for the SC phase transition when $\xi_{\bm k}$ is p-h symmetric.
\par
The Higgs mode is uncoupled with the single-particle states $\ket{e_{\bm k_F}}$ due to their opposite parity under $\mathcal C_{\bm k_F}$ despite their degeneracy at $2\Delta_0$. 
The Higgs mode has even parity for all $\mathcal C_{\bm k\ulin{\bm k}}$ and $\mathcal C_{\bm k_F}$ because $[\mathcal C_{\bm k\ulin{\bm k}},\beta_{\rm H}^\dagger]=[\mathcal C_{\bm k_F},\beta_{\rm H}^\dagger]=0$. A single Higgs mode $\beta_{\rm H}^\dagger\ket{\Psi}$ in fact consists of the single-particle states $\ket{e_{\bm k\ulin{\bm k}}^0}$ \cite{supplement}.
Thus, the Higgs mode is a stable excitation of $\mathcal H_I$. 

{\it Strong-coupling perturbation theory.}-- 
\begin{figure}
\centering
\includegraphics[width=\linewidth]{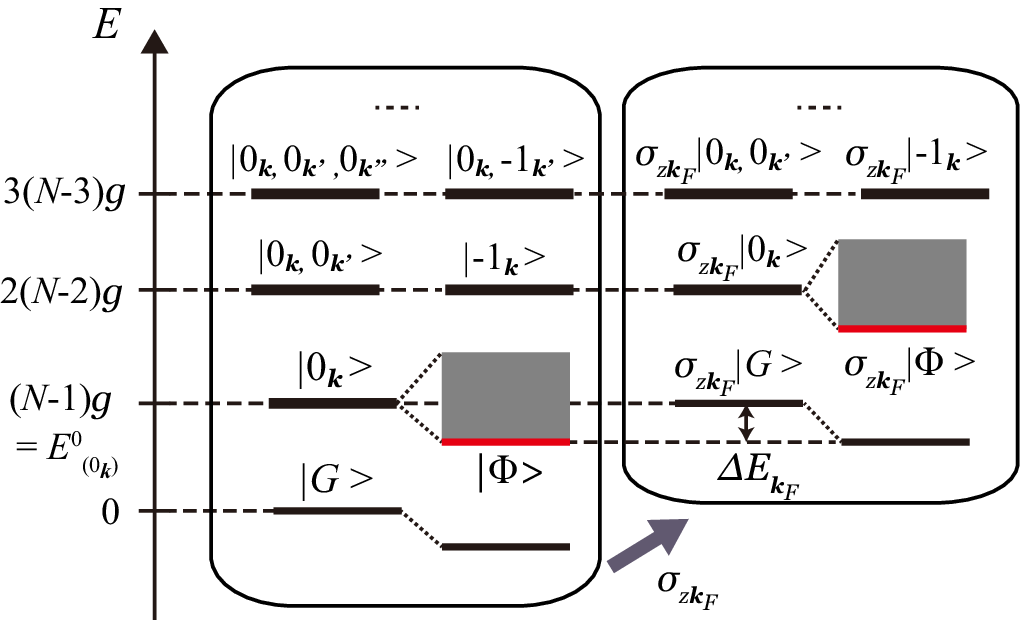}
\caption{Low-lying energy eigenstates of $\mathcal H_X$ and their energy shifts by $\mathcal H_K$. We denote $\ket{m_{\bm k},n_{\bm k'},\cdots}\equiv\ket{m_x}_{\bm k\ulin{\bm k}}\otimes\ket{n_x}_{\bm k'\ulin{\bm k'}}\otimes\cdots \otimes\prod'_{\bm k''\neq \bm k,\bm k'}\ket{1_x}_{\bm k''\ulin{\bm k}''}\otimes\prod_{\bm k_F}\ket{+}_{\bm k_F}$ ($m,n=0,-1$). The eigenstates of $\mathcal H_X$ can be decomposed into the blocks represented by squares, where each block is characterized by the parity under $\mathcal C_{\bm k_F}$. The multiple Higgs modes $\ket{\Phi}$ and $\sigma_{z\bm k_F}\ket{\Phi}$ are the bound states formed at the lower edges of the continuum in each block shown as the gray regions.}
\label{fig.spectrum}
\end{figure}
Having shown the emergence of the Higgs mode due to breakdown of $\mathbb Z_2$ in the weak-coupling theory, we demonstrate it as well in a strong-coupling approach, where we treat $\mathcal H_K$ as a perturbation to the unperturbed Hamiltonian $\mathcal H_X$. It is valid when $Ng$ is much larger than the band width, i.e., $Ng\gg {\rm max}|\xi_{\bm k}|$, where $N$ is the total number of wave vectors.
Given that the Higgs mode has even parity for all $\mathcal C_{\bm k\ulin{\bm k}}$, we treat $\bm S_{\bm k\ulin{\bm k}}$ as a spin-1 operator to take into account only even parity states under $\mathcal C_{\bm k\ulin{\bm k}}$.
\par
Each eigenstate of $\mathcal H_{\rm MF}$ has a corresponding eigenstate of $\mathcal H_X$, where the former transforms to the latter in the strong-coupling limit. 
$\mathbb Z_2$ under $\mathcal T$ is broken in the unperturbed ground states $\ket{G}=\prod'_{\bm k}\ket{1_x}_{\bm k\ulin{\bm k}}\otimes\prod_{\bm k_F}\ket{+}_{\bm k_F}$ and $\ket{\bar G}= \mathcal T\ket{G}$, where $\ket{m_x}_{\bm k\ulin{\bm k}}$ ($m=1,0,-1$) denote the spin-1 triplet states for $S_{x\bm k\ulin{\bm k}}$ \cite{supplement}. They have gap functions $\Delta=Ng/2\equiv\tilde\Delta_0$ and $\Delta=-\tilde\Delta_0$, respectively.
We consider perturbation to $\ket{G}$ and the low-lying states above it shown in Fig.~\ref{fig.spectrum}.  
\par
The energy eigenstates of $\mathcal H_X$ can be decomposed into the blocks characterized by the parity under $\mathcal C_{\bm k_F}$, which are represented as the squares in Fig.~\ref{fig.spectrum}. $\mathcal H_K$ has no matrix elements between states in different blocks. 
If we focus on a block that has odd parity under $\mathcal C_{\bm k_F}$ for a set of Fermi wavevectors $\{\bm k_F\}$, each unperturbed state in this block can be obtained by operating $\mathcal Z(\{\bm k_F\})\equiv\prod_{\bm k_F'\in\{\bm k_F\}}\sigma_{z\bm k_F'}$ on a corresponding state in the block of $\ket{G}$.
Flipping a pseudospin on the FS from $\ket{+}_{\bm k_F}$ to $\ket{-}_{\bm k_F}$ by $\sigma_{z\bm k_F}$ yields a higher excited state as
\begin{eqnarray}
\mathcal H_X(\sigma_{z\bm k_F}\ket{\phi})=(-gS_x(\phi)^2+\Delta E_{\phi})(\sigma_{z\bm k_F}\ket{\phi})~,\label{eq.sigmaz}
\end{eqnarray}
where $\mathcal C_{\bm k_F}\ket{\phi}=\ket{\phi}$, $S_x\ket{\phi}=S_x(\phi)\ket{\phi}$, and $\Delta E_\phi=g(2S_x(\phi)-1)$. 
The energy spectrum of $\mathcal H_X$ in Fig.~\ref{fig.spectrum} is obtained by using Eq.~(\ref{eq.sigmaz}).
\par
The degeneracy among the first-excited states $\{\ket{0_{\bm k}}\}$ is lifted by diagonalizing the second-order effective Hamiltonian $\mathcal H_{\rm eff}\equiv\mathcal H_K(E^0_{(0_{\bm k})}-\mathcal H_X)^{-1}\mathcal H_K$, where $E^0_{(0_{\bm k})}\equiv (N-1)g$ is the unperturbed energy of $\ket{0_{\bm k}}$.
The effective Hamiltonian can be written as
\begin{eqnarray}
\mathcal H_{\rm eff}=\sideset{}{^{'}}\sum_{\bm k}J_{\bm k}\ket{0_{\bm k}}\bra{0_{\bm k}}-\sideset{}{^{'}}\sum_{\bm k\neq\bm k' }J_{\bm k,\bm k'}\ket{0_{\bm k}}\bra{ 0_{\bm k'}}~,\label{eq.Heff}
\end{eqnarray}
where
\begin{eqnarray}
J_{\bm k}=\frac{2\xi_{\bm k}^2}{(N-1)g},\quad J_{\bm k,\bm k'}=\frac{4\xi_{\bm k}\xi_{\bm k'}}{(N-1)(N-3)g}.
\end{eqnarray}
Note that energy in Eq.~(\ref{eq.Heff}) is measured from $E^0_{(0_{\bm k})}-\Delta E_{\bm k_F}$, where $\Delta E_{\bm k_F}=\frac{2\sum_{\bm k'}'\xi_{\bm k'}^2}{(N-3)g}$ is the second-order energy shift of $\sigma_{z\bm k_F}\ket{G}$ that corresponds to $\ket{e_{\bm k_F}}$.
The first term in Eq.~(\ref{eq.Heff}) represents the energy continuum above the threshold $E^0_{(0_{\bm k})}-\Delta E_{\bm k_F}$, as shown in Fig.~\ref{fig.spectrum}. 
\par
We consider the following state:
\begin{equation}
\ket{\Phi}=\sideset{}{^{'}}\sum_{\bm k}\frac{1}{\xi_{\bm k}}\ket{0_{\bm k}}.
\end{equation}
Using Eq.~(\ref{eq.Heff}), we obtain
\begin{eqnarray}
\mathcal H_{\rm eff}\ket{\Phi}=\sideset{}{^{'}}\sum_{\bm k}C_{\bm k}\ket{0_{\bm k}},~C_{\bm k}=\frac{J_{\bm k}}{\xi_{\bm k}}+\sum_{\bm k'\neq \bm k}\left(-\frac{J_{\bm k\bm k'}}{\xi_{\bm k'}}\right).\label{eq.bdgenergy}
\end{eqnarray}
In the limit $N\gg N_{k_F}$, using $\sum_{\bm k'\neq \bm k}'1\simeq N/2$, Eq.~(\ref{eq.bdgenergy}) reduces to $\mathcal H_{\rm eff}|\Phi\rangle=0$. Remarkably, $|\Phi\rangle$ represents a bound state of $\{\ket{0_{\bm k}}\}$ that is formed at the lower edge of the energy continuum at $2\tilde{\Delta}_0$. 
On the other hand, we find $\beta_{\rm H}^\dagger|\Psi\rangle\to |\Phi\rangle$ in the strong-coupling limit using $\beta_{\rm H}^\dagger\to -i\sum_{\bm k}'(S_{y\bm k\ulin{\bm k}}-iS_{z\bm k\ulin{\bm k}})/\xi_{\bm k}$. 
Thus, the Higgs mode $\ket{\Phi}$ arises in the strong-coupling theory.
$\ket{\Phi}$ is indeed stable, because it is uncoupled with $\sigma_{z\bm k_F}\ket{G}$.
In the weak-coupling regime, since $\ket{0_{\bm k}}$ corresponds to $\ket{e^0_{\bm k\ulin{\bm k}}}$, $\beta_{\rm H}^\dagger\ket{\Psi}$ is considered a bound state of $\{\ket{e^0_{\bm k\ulin{\bm k}}}\}$ formed at $2\Delta_0$. 
\par
{\it Multiple Higgs modes.}--
We expect that a bound state analogous to the Higgs mode may be formed among the first excited states in each block.
The effective Hamiltonian $\mathcal H^{(n)}_{\rm eff}$ for the degenerate first excited states $\{\mathcal Z(\{\bm k_F\})\ket{0_{\bm k}}\}$ in the block of $\mathcal Z(\{\bm k_F\})\ket{G}$ is given by
\begin{eqnarray}
\mathcal H_{\rm eff}^{(n)}=\sideset{}{^{'}}\sum_{\bm k}J^{(n)}_{\bm k}\mathcal Z\ket{0_{\bm k}}\bra{ 0_{\bm k}}\mathcal Z-\sideset{}{^{'}}\sum_{\bm k\neq\bm k' }J^{(n)}_{\bm k,\bm k'}\mathcal Z\ket{0_{\bm k}}\bra{ 0_{\bm k'}}\mathcal Z~,\label{eq.Heffn}
\end{eqnarray}
where
\begin{eqnarray}
&&J^{(n)}_{\bm k}=\frac{2\xi_{\bm k}^2}{(N-(2n+1))g}~,\\
&&J^{(n)}_{\bm k,\bm k'}=\frac{4\xi_{\bm k}\xi_{\bm k'}}{(N-(2n+1))(N-(2n+3))g}~.
\end{eqnarray}
Here, $n$ is the number of wave vectors in $\{\bm k_F\}$. 
In the limit $N\gg N_{k_F}$, we find that $\mathcal H^{(n)}_{\rm eff}$ has a Higgs-like bound state
\begin{eqnarray}
\ket{\Phi^{(n)}}=\mathcal Z(\{\bm k_F\})\ket{\Phi}~.
\end{eqnarray}
Thus, combining with the original Higgs mode $\ket{\Phi^{(0)}}=\ket{\Phi}$, the multiple Higgs modes $\ket{\Phi^{(n)}}$ ($0\leq n\leq N_{k_F}$) have quantized energy $2(n+1)\tilde\Delta_0$. They have odd parity under $\mathcal C_{\bm k_F}$ for $n$ Fermi wavevectors in $\{\bm k_F\}$. They should appear in the weak-coupling regime. They correspond to the states $\mathcal Z(\{\bm k_F\})\beta_{\rm H}^\dagger\ket{\Psi}$ with quantized energy $2(n+1)\Delta_0$ in the weak-coupling regime.
\par
These multiple Higgs modes may be observable in superconductors with CDW-order such as NbSe$_2$ using Raman spectroscopy  \cite{sooryakumar-81}. Their signature may appear as multiple resonance peaks at frequencies $2(n+1)\Delta_0$ in Raman spectrum. 
It may be also possible to observe them using recently developed terahertz spectroscopy techniques \cite{matsunaga-13}.
\par
{\it Conclusions and conjectures.}--
We have shown that the Higgs mode emerges due to spontaneous breakdown of $\mathbb Z_2$ under $\mathcal T$ in a superconductor. We further predicted the existence of the multiple Higgs modes that have quantized energy $2(n+1)\Delta_0$ ($0\leq n\leq N_{k_F}$).
\par
The analysis in the present work has broad applicability in condensed-matter and particle physics. Multiple Higgs modes may appear in systems for which the BCS theory is applicable, such as fermionic superfluids, atomic nuclei, and quark matters. It would be reasonable to conjecture that Higgs modes in other condensed-matter systems, such as a bosonic superfluid in an optical lattice and quantum spin systems, also arise due to spontaneous breakdown of $\mathbb Z_2$ symmetries. Extensions of the present work to these systems are left for the future. 

\begin{acknowledgements}
The author wishes to thank I. Danshita, D. Yamamoto, and R. Yoshii for fruitful discussions. This work is supported by the Japan Society for the Promotion of Science Grant-in-Aid for Scientific Research (KAKENHI Grant No.~19K03691).
\end{acknowledgements}


\appendix

\section{Supplemental Materials}
\subsection{Pseudospin Operators}

In this Section, we summarize the algebra for the pseudospin operators.
We introduce $\bm S_{\bm k\ulin{\bm k}}$ in Eq.~(9) in the main text. We also introduce another pseudospin operator $\tilde{\bm S}_{\bm k\ulin{\bm k}}$ as
\begin{eqnarray}
\tilde S_{\mu\bm k\ulin{\bm k}}=S_{\mu\bm k}+(-1)^{\delta_{\mu,x}}S_{\mu\ulin{\bm k}}~,~(\mu=x,y,z),
\end{eqnarray}
where $\bm k$ is above the Fermi surface (FS) ($\xi_{\bm k}>0$). 
$\bm S_{\bm k}$ is transformed by $\mathcal C_{\bm k\ulin{\bm k}}$ and $\mathcal C_{\bm k_F}$ as
\begin{eqnarray}
\mathcal C_{\bm k\ulin{\bm k}}S_{\mu\bm k'}\mathcal C_{\bm k\ulin{\bm k}}=
\left\{
\begin{array}{ll}
(-1)^{\delta_{\mu,x}+1}S_{\mu\ulin{\bm k}'}~, & (\bm k'=\bm k,\ulin{\bm k})~, \\
S_{\mu{\bm k}'}~, & {\rm otherwise},
\end{array}
\right.
\label{eq.Ck}\\
\mathcal C_{\bm k_F}S_{\mu\bm k}\mathcal C_{\bm k_F}=
\left\{
\begin{array}{ll}
(-1)^{\delta_{\mu,x}+1}S_{\mu\bm k_F}~, & (\bm k=\bm k_F)~, \\
S_{\mu\bm k}~, & {\rm otherwise}.
\end{array}
\right.\label{eq.CkF}
\end{eqnarray}
Using Eqs.~(\ref{eq.Ck}) and (\ref{eq.CkF}), one can derive the commutation relations
\begin{eqnarray}
[\mathcal C_{\bm k\ulin{\bm k}},S_{\mu\bm k\ulin{\bm k}}]=\{\mathcal C_{\bm k\ulin{\bm k}},\tilde S_{\mu\bm k\ulin{\bm k}}\}=0~.
\end{eqnarray}
${\bm S}_{\bm k\ulin{\bm k}}$ and $\tilde{\bm S}_{\bm k\ulin{\bm k}}$ also satisfy
\begin{eqnarray}
&&[S_{\mu\bm k\underline{\bm k}},S_{\nu\bm k\underline{\bm k}}]=[\tilde S_{\mu\bm k\underline{\bm k}},\tilde S_{\nu\bm k\underline{\bm k}}]=i\sum_\rho \varepsilon_{\mu\nu\rho} S_{\rho\bm k\underline{\bm k}}~,\label{eq.commuSS}\\
&&[S_{\mu\bm k\ulin{\bm k}},\tilde S_{\nu\bm k\ulin{\bm k}}]=i\sum_\rho \varepsilon_{\mu\nu\rho}\tilde S_{\rho\bm k\underline{\bm k}}~,\label{eq.commuSSt}\\
&&S_{\mu\bm k\ulin{\bm k}}\tilde S_{\mu\bm k\ulin{\bm k}}=\tilde S_{\mu\bm k\ulin{\bm k}}S_{\mu\bm k\ulin{\bm k}}=0~,\quad S_{\mu\bm k\ulin{\bm k}}^2+\tilde S_{\mu\bm k\ulin{\bm k}}^2=1~.
\end{eqnarray}
\par
The spin-1 triplet states for $S_{z\bm k\ulin{\bm k}}$ can be written as
\begin{eqnarray}
&&|1\rangle_{\bm k\underline{\bm k}}=|\upa\dna\rangle_{\bm k\underline{\bm k}},\\
&&|0\rangle_{\bm k\underline{\bm k}}=\frac{1}{\sqrt{2}}(|\upa\upa\rangle_{\bm k\underline{\bm k}}+|\dna\dna\rangle_{\bm k\underline{\bm k}}),\\
&&|-1\rangle_{\bm k\underline{\bm k}}=|\dna\upa\rangle_{\bm k\underline{\bm k}}~,
\end{eqnarray}
Here, we denote $\ket{s~t}_{\bm k\underline{\bm k}}=\ket{s}_{\bm k}\otimes \ket{t}_{\ulin{\bm k}}$ ($s,t=\upa,\dna$). 
They satisfy $S_{z\bm k\underline{\bm k}}|m\rangle_{\bm k\underline{\bm k}} =m|m\rangle_{\bm k\underline{\bm k}}$ ($m=1,0,-1$) and have even parity under $\mathcal C_{\bm k\ulin{\bm k}}$: $\mathcal C_{\bm k\ulin{\bm k}}|m\rangle_{\bm k\underline{\bm k}}=|m\rangle_{\bm k\underline{\bm k}}$. 
\par
The spin-1 triplet states for $S_{x\bm k\ulin{\bm k}}$ can be written as
\begin{eqnarray}
\ket{1_x}_{\bm k\underline{\bm k}}&=&\ket{++}_{\bm k\ulin{\bm k}}~,\\
\ket{0_x}_{\bm k\underline{\bm k}}&=&\frac{1}{\sqrt{2}}(\ket{-+}_{\bm k\ulin{\bm k}}-\ket{+-}_{\bm k\ulin{\bm k}})~,\\
\ket{-1_x}_{\bm k\underline{\bm k}}&=&-\ket{--}_{\bm k\ulin{\bm k}}~,
\end{eqnarray}
where $\ket{\pm}_{\bm k}=(\ket{\upa}_{\bm k}\pm\ket{\dna}_{\bm k})/\sqrt{2}$. They satisfy $S_{x\bm k\ulin{\bm k}}|m_x\rangle_{\bm k\ulin{\bm k}}=m|m_x\rangle_{\bm k\ulin{\bm k}}$ ($m=1,0,-1$) and have even parity under $\mathcal C_{\bm k\ulin{\bm k}}$: $\mathcal C_{\bm k\ulin{\bm k}}|m_x\rangle_{\bm k\underline{\bm k}}=|m_x\rangle_{\bm k\underline{\bm k}}$.
\par
The spin-0 singlet state $|\tilde 0\rangle_{\bm k\ulin{\bm k}}$ can be written as
\begin{eqnarray}
|\tilde 0\rangle_{\bm k\ulin{\bm k}}&=&\frac{1}{\sqrt{2}}(\ket{\upa\upa}_{\bm k\ulin{\bm k}}-\ket{\dna\dna}_{\bm k\ulin{\bm k}})\\
&=&\frac{1}{\sqrt{2}}(\ket{+-}_{\bm k\ulin{\bm k}}+\ket{-+}_{\bm k\ulin{\bm k}}).
\end{eqnarray}
It satisfies $S_{\mu\bm k\ulin{\bm k}}|\tilde 0\rangle_{\bm k\ulin{\bm k}}=0$ ($\mu=x,y,z$) and has odd parity under $\mathcal C_{\bm k\ulin{\bm k}}$: $\mathcal C_{\bm k\ulin{\bm k}}|\tilde 0\rangle_{\bm k\ulin{\bm k}}=-|\tilde 0\rangle_{\bm k\ulin{\bm k}}$. We note that the pseudospins for $\bm k$ and $\ulin{\bm k}$ are entangled in $|0\rangle_{\bm k\ulin{\bm k}}$, $|0_x\rangle_{\bm k\ulin{\bm k}}$, and $|\tilde{0}\rangle_{\bm k\ulin{\bm k}}$.
\par
$S^\pm_{\bm k\ulin{\bm k}}=S_{x\bm k\ulin{\bm k}}\pm iS_{y\bm k\ulin{\bm k}}$ are the ladder operators for the spin-1 triplet states $\{\ket{m}_{\bm k\ulin{\bm k}}\}$:
\begin{eqnarray}
&&S^-_{\bm k\ulin{\bm k}}\ket{1}_{\bm k\ulin{\bm k}}=S^+_{\bm k\ulin{\bm k}}\ket{-1}_{\bm k\ulin{\bm k}}=\sqrt{2}\ket{0}_{\bm k\ulin{\bm k}}~,\label{eq.ladderop1}\\
&&S^+_{\bm k\ulin{\bm k}}\ket{1}_{\bm k\ulin{\bm k}}=S^-_{\bm k\ulin{\bm k}}\ket{-1}_{\bm k\ulin{\bm k}}=0~,\label{eq.ladderop2}\\
&&S^\pm_{\bm k\ulin{\bm k}}\ket{0}_{\bm k\ulin{\bm k}}=
\left\{
\begin{array}{l}
\sqrt{2}\ket{1}_{\bm k\ulin{\bm k}}~,\\
\sqrt{2}\ket{-1}_{\bm k\ulin{\bm k}}~.
\end{array}
\right.\label{eq.ladderop3}
\end{eqnarray}
$\tilde{S}^\pm_{\bm k\ulin{\bm k}}=\tilde{S}_{x\bm k\ulin{\bm k}}\pm i\tilde{S}_{y\bm k\ulin{\bm k}}$ transform the spin-1 triplet states into the spin-0 singlet state $\ket{\tilde{0}}_{\bm k\ulin{\bm k}}$ and {\it vice versa} as
\begin{eqnarray}
&&\tilde{S}^-_{\bm k\ulin{\bm k}}\ket{1}_{\bm k\ulin{\bm k}}=-\tilde{S}^+_{\bm k\ulin{\bm k}}\ket{-1}_{\bm k\ulin{\bm k}}=-\sqrt{2}\ket{\tilde{0}}_{\bm k\ulin{\bm k}}~,\label{eq.Stil1}\\
&&\tilde{S}^+_{\bm k\ulin{\bm k}}\ket{1}_{\bm k\ulin{\bm k}}=\tilde{S}^-_{\bm k\ulin{\bm k}}\ket{-1}_{\bm k\ulin{\bm k}}=\tilde{S}^\pm_{\bm k\ulin{\bm k}}\ket{0}_{\bm k\ulin{\bm k}}=0~,\label{eq.Stil2}\\
&&\tilde{S}^+_{\bm k\ulin{\bm k}}\ket{\tilde{0}}_{\bm k\ulin{\bm k}}=-\sqrt{2}\ket{1}_{\bm k\ulin{\bm k}},\\
&&\tilde{S}^-_{\bm k\ulin{\bm k}}\ket{\tilde{0}}_{\bm k\ulin{\bm k}}=\sqrt{2}\ket{-1}_{\bm k\ulin{\bm k}}~.\label{eq.Stil3}
\end{eqnarray}
\par
In general, operating $\tilde S_{\mu\bm k\ulin{\bm k}}$ on a certain state changes its parity under $\mathcal C_{\bm k\ulin{\bm k}}$, while operating $S_{\mu\bm k\ulin{\bm k}}$ does not as
\begin{eqnarray}
\mathcal C_{\bm k\ulin{\bm k}}S_{\mu\bm k\ulin{\bm k}}\ket{\psi_{\rm e(o)}}_{\bm k\ulin{\bm k}}
&=&\pm S_{\mu\bm k\ulin{\bm k}}\ket{\psi_{\rm e(\rm o)}}_{\bm k\ulin{\bm k}},~\label{eq.CS}\\
\mathcal C_{\bm k\ulin{\bm k}}\tilde S_{\mu\bm k\ulin{\bm k}}\ket{\psi_{\rm e(\rm o)}}_{\bm k\ulin{\bm k}}
&=&\mp \tilde S_{\mu\bm k\ulin{\bm k}}\ket{\psi_{\rm e(\rm o)}}_{\bm k\ulin{\bm k}}~,\label{eq.CStilde}
\end{eqnarray}
where $\ket{\psi_{\rm e}}_{\bm k\ulin{\bm k}}$ ($\ket{\psi_{\rm o}}_{\bm k\ulin{\bm k}}$) denotes an even (odd) parity state under $\mathcal C_{\bm k\ulin{\bm k}}$. The upper (lower) signs on the right-hand-sides of Eqs.~(\ref{eq.CS}) and (\ref{eq.CStilde}) are for the even (odd) parity state.

\subsection{Applications of Pseudospin Operators}

In this Section, for illustration of applications of the pseudospin operators introduced in the previous Section, we describe the normal state in terms of them and also derive some rigorous relations for occupation numbers of fermions and dynamical spin susceptibilities.
\par
We first consider the normal state. Given the kinetic energy term in Eq.~(10), the ground state of $\mathcal H_K$ can be written as $\ket{\Psi_0}=\prod_{\bm k}'\ket{-1}_{\bm k\ulin{\bm k}}\otimes \ket{\Phi_F}$, where $\ket{\Phi_F}$ denotes the wave function for the pseudospins on the FS, which are zero modes for $\mathcal H_K$.
Using Eqs.~(\ref{eq.commuSS}) and (\ref{eq.commuSSt}), we obtain
\begin{eqnarray}
[\mathcal H_K,S_{\bm k\ulin{\bm k}}^\pm]&=&\pm 2\xi_{\bm k}S_{\bm k\ulin{\bm k}}^\pm~,\label{eq.commuH_K1}\\
\protect[\mathcal H_K,\tilde S_{\bm k\ulin{\bm k}}^\pm]&=&\pm 2\xi_{\bm k}\tilde S_{\bm k\ulin{\bm k}}^\pm~.\label{eq.commuH_K2}
\end{eqnarray}
The above equations show that $S_{\bm k\ulin{\bm k}}^\pm$ and $\tilde S_{\bm k\ulin{\bm k}}^\pm$ are the raising and lowering operators for the eigenstates of $\mathcal H_K$. Thus, $S_{\bm k\ulin{\bm k}}^+\ket{\Psi_0}/\sqrt{2}$ and $\tilde S_{\bm k\ulin{\bm k}}^+\ket{\Psi_0}/\sqrt{2}$ are the degenerate excited states with excitation energy $2\xi_{\bm k}$ involving a pair of fermions, where they have even and odd parity under $\mathcal C_{\bm k\ulin{\bm k}}$, respectively.
\par
We derive the fermion occupation number for $\ket{\psi_{\rm e(o)}}_{\bm k\ulin{\bm k}}$. Using Eq.~(\ref{eq.CStilde}), we obtain 
\begin{equation}
_{\bm k\ulin{\bm k}}\bra{\psi_{\rm e(o)}}\tilde {\bm S}_{\bm k\ulin{\bm k}}\ket{\psi_{\rm e(o)}}_{\bm k\ulin{\bm k}}=0~.\label{eq.Skave}
\end{equation}
Specifically, from the $z$-component of Eq.~(\ref{eq.Skave}) we obtain the quantization of fermion occupation number 
\begin{equation}
_{\bm k\ulin{\bm k}}\bra{\psi_{\rm e(o)}}(n_{\bm k\uparrow}+n_{-\bm k\downarrow}+n_{\ulin{\bm k}\uparrow}+n_{-\ulin{\bm k}\downarrow})\ket{\psi_{\rm e(o)}}_{\bm k\ulin{\bm k}}=2~,
\end{equation}
where $n_{\bm k\sigma}=c_{\bm k\sigma}^\dagger c_{\bm k\sigma}$ is the fermion number operator.
We also obtain
\begin{equation}
_{\bm k_F}\bra{\pm}\bm S_{\bm k_F}\ket{\pm}_{\bm k_F}=(\pm 1/2,0,0)~,\label{eq.Skfave}
\end{equation}
where $\ket{+}_{\bm k_F}$ ($\ket{-}_{\bm k_F}$) is an even (odd) parity state under $\mathcal C_{\bm k_F}$. The upper (lower) sign on the right-hand-side of Eq.~(\ref{eq.Skfave}) is for $\ket{+}_{\bm k_F}$ ($\ket{-}_{\bm k_F}$).
From the $z$-component of Eq.~(\ref{eq.Skfave}), we obtain 
\begin{equation}
_{\bm k_F}\bra{\pm}(n_{\bm k_F\uparrow}+n_{-\bm k_F\downarrow})\ket{\pm}_{\bm k_F}=1~.
\end{equation}
Therefore, the states $\ket{\psi_{\rm e(o)}}_{\bm k\ulin{\bm k}}$ and $\ket{\pm}_{\bm k_F}$ that appear in the eigenstates of $\mathcal H_I$ are half-filled.
\par
One can also derive some rigorous relations for the dynamical spin susceptibility
\begin{equation}
\chi_{\mu\bm k,\nu}(\omega)\equiv-i\int_0^\infty\langle[S_\nu,S_{\mu\bm k}(t)]\rangle e^{-i\omega t}dt~,
\end{equation}
where $S_{\mu\bm k}(t)=e^{-i\mathcal H_It}S_{\mu\bm k}e^{i\mathcal H_It}$ is the Heisenberg representation and $\langle\cdots\rangle$ denotes the thermal average. We focus on $\chi_{y\bm k,x}(\omega)$ and evaluate
\begin{equation} 
\langle[S_x,S_{y\bm k}(t)]\rangle\propto \sum_n e^{-\beta E_n}\bra{n}[S_x,S_{y\bm k}(t)]\ket{n}~,
\end{equation}
where $\beta$ is the inverse temperature and $\ket{n}$ denotes an eigenstate of $\mathcal H_I$: $\mathcal H_I\ket{n}=E_n\ket{n}$. Since operating $\tilde S_{y\bm k\ulin{\bm k}}$ changes the parity of the state under $\mathcal C_{\bm k\ulin{\bm k}}$, we obtain $\bra{n}S_x\tilde S_{y\bm k\ulin{\bm k}}(t)\ket{n}=\bra{n}\tilde S_{y\bm k\ulin{\bm k}}(t)S_x\ket{n}=0$ and therefore $\bra{n}S_xS_{y\bm k}(t)\ket{n}=-\bra{n}S_xS_{y\ulin{\bm k}}(t)\ket{n}$ and $\bra{n}S_{y\bm k}(t)S_x\ket{n}=-\bra{n}S_{y\ulin{\bm k}}(t)S_x\ket{n}$. On the other hand, one can also derive $\bra{n}S_xS_{y\bm k_F}(t)\ket{n}=0$ using $\mathcal C_{\bm k_F}S_{y\bm k_F}(t)\mathcal C_{\bm k_F}=-S_{y\bm k_F}(t)$.
We thus obtain
\begin{eqnarray}
&&\chi_{y\bm k,x}(\omega)=-\chi_{y\ulin{\bm k},x}(\omega)~,\label{eq.chiykx}\\
&&\chi_{y\bm k_F,x}(\omega)=0~.\label{eq.chiykFx}
\end{eqnarray}
Equations~(\ref{eq.chiykx}) and (\ref{eq.chiykFx}) lead to the out-of-phase oscillations of spin fluctuations for the Higgs mode $\delta S_{y\ulin{\bm k}}=-\delta S_{y\bm k}$ in the classical spin analysis [42].
Analogously, one can show
\begin{eqnarray}
&&\chi_{x\bm k,y}(\omega)=-\chi_{x\ulin{\bm k},y}(\omega)~,\label{eq.chixky}\\
&&\chi_{x\bm k_F,y}(\omega)=0~,\label{eq.chixkFy}
\end{eqnarray}
using $\bra{n}\tilde S_{y\bm k\ulin{\bm k}}S_{x\bm k\ulin{\bm k}}(t)\ket{n}=\bra{n}S_{x\bm k\ulin{\bm k}}(t)\tilde S_{y\bm k\ulin{\bm k}}\ket{n}=0$, $\bra{n}S_{y\bm k_F}S_{x\bm k\ulin{\bm k}}(t)\ket{n}=\bra{n}S_{x\bm k\ulin{\bm k}}(t)S_{y\bm k_F}\ket{n}=0$, and $\bra{n}S_yS_{x\bm k\ulin{\bm k}}(t)\ket{n}=\bra{n}S_{x\bm k\ulin{\bm k}}(t)S_y\ket{n}=0$. We thus find that the dynamical spin susceptibilities for total spin vanish as $\chi_{xy}(\omega)=\sum_{\bm k}\chi_{x\bm k,y}(\omega)=0$ and $\chi_{yx}(\omega)=\sum_{\bm k}\chi_{y\bm k,x}(\omega)=0$ [42].
The expressions for the MF theory 
\begin{equation}
\chi_{y\bm k,x}(\omega)\propto \chi_{x\bm k,y}(\omega)\propto \frac{\omega\xi_{\bm k}}{E_{\bm k}(4E_{\bm k}^2-\omega^2)}~,
\end{equation}
indeed satisfy Eqs.~(\ref{eq.chiykx})$\sim$(\ref{eq.chixkFy}) [42].

\subsection{Pseudospin Operators for Bogolons}

In this Section, we summarize the algebra for the pseudospin operators for bogolons. $\bm S'_{\bm k}$ is introduced as [42]
\begin{eqnarray}
\left(
\begin{array}{ccc}
S_{z\bm k}'\\
S_{x\bm k}'\\
S_{y\bm k}'
\end{array}
\right)=
\left(
\begin{array}{ccc}
-\cos\varphi_{\bm k} & -\sin\varphi_{\bm k} & 0\\
\sin\varphi_{\bm k} & -\cos\varphi_{\bm k} & 0\\
0 & 0 & 1
\end{array}
\right)
\left(
\begin{array}{cc}
S_{z\bm k}\\
S_{x\bm k}\\
S_{y\bm k}
\end{array}
\right),\label{eq.Sprime}
\end{eqnarray}
where $\cos\varphi_{\bm k}=-\xi_{\bm k}/E_{\bm k}$ and $\sin\varphi_{\bm k}=\Delta_0/E_{\bm k}$.
$\bm S_{\bm k}$ is rotated about the angle $\pi-\varphi_{\bm k}$ in the $xz$-plane in Eq.~(\ref{eq.Sprime}).
Using Eq.~(\ref{eq.Sprime}), the MF Hamiltonian can be written in the form of Eq.~(11).
The eigenstates of $S_{z\bm k}'$ are given as
\begin{eqnarray}
|\uparrow'\rangle_{\bm k}&=&u_{\bm k}|\uparrow\rangle_{\bm k}-v_{\bm k}|\downarrow\rangle_{\bm k}~,\\
|\!\downarrow'\rangle_{\bm k}&=&u_{\bm k}|\downarrow\rangle_{\bm k}+v_{\bm k}|\uparrow\rangle_{\bm k}~,
\end{eqnarray}
where $\ket{\dna'}_{\bm k}$ represents the vacuum of bogolons and $\ket{\upa'}_{\bm k}$ the excited state with a pair of bogolons. All the pseudospins ${\bm S}_{\bm k}'$ are aligned downward in the $z$ direction in the superconducting (SC) ground state. The BCS wave function can be thus written as
\begin{equation}
|\Psi\rangle=\prod_{\bm k}\ket{\dna'}_{\bm k}.
\end{equation}
Note that $(S'_{x\bm k_F},S'_{y\bm k_F},S'_{z\bm k_F})=(S_{z\bm k_F},S_{y\bm k_F},-S_{x\bm k_F})$, $\ket{\upa'}_{\bm k_F}=\ket{-}_{\bm k_F}$, and $\ket{\dna'}_{\bm k_F}=\ket{+}_{\bm k_F}$.
\par
We introduce another pseudospin operator $\tilde{\bm S}'_{\bm k\ulin{\bm k}}$ as
\begin{eqnarray}
\tilde S'_{\mu\bm k\ulin{\bm k}}=S'_{\mu\bm k}+(-1)^{\delta_{\mu,z}}S'_{\mu\ulin{\bm k}}~,~(\mu=x,y,z),
\end{eqnarray}
where $\bm k$ is above the FS ($\xi_{\bm k}>0$).
$\bm S'_{\bm k}$ is transformed by $\mathcal C_{\bm k\ulin{\bm k}}$ and $\mathcal C_{\bm k_F}$ as
\begin{eqnarray}
\mathcal C_{\bm k\ulin{\bm k}}S'_{\mu\bm k'}\mathcal C_{\bm k\ulin{\bm k}}=
\left\{
\begin{array}{ll}
(-1)^{\delta_{\mu,z}+1}S'_{\mu\ulin{\bm k}'}~, & (\bm k'=\bm k,\ulin{\bm k})~, \\
S'_{\mu{\bm k}'}~, & {\rm otherwise},
\end{array}
\right.
\label{eq.CkSd}\\
\mathcal C_{\bm k_F}S'_{\mu\bm k}\mathcal C_{\bm k_F}=
\left\{
\begin{array}{ll}
(-1)^{\delta_{\mu,z}+1}S'_{\mu\bm k_F}~, & (\bm k=\bm k_F)~, \\
S_{\mu\bm k}~, & {\rm otherwise}.
\end{array}
\right.\label{eq.CkFSd}
\end{eqnarray}
Using Eqs.~(\ref{eq.CkSd}) and (\ref{eq.CkFSd}), one can derive the commutation relations
\begin{eqnarray}
[\mathcal C_{\bm k\ulin{\bm k}},S'_{\mu\bm k\ulin{\bm k}}]=\{\mathcal C_{\bm k\ulin{\bm k}},{\tilde S}'_{\mu\bm k\ulin{\bm k}}\}=0~.
\end{eqnarray}
$\bm S'_{\bm k\ulin{\bm k}}$ and $\tilde{\bm S}'_{\bm k\ulin{\bm k}}$ obey the commutation relations
\begin{eqnarray}
&&[S'_{\mu\bm k\underline{\bm k}},S'_{\nu\bm k\underline{\bm k}}]=[\tilde S'_{\mu\bm k\underline{\bm k}},\tilde S'_{\nu\bm k\underline{\bm k}}]=i\sum_\rho \varepsilon_{\mu\nu\rho} S'_{\rho\bm k\underline{\bm k}}~,\label{eq.commuSSd}\\
&&[S'_{\mu\bm k\ulin{\bm k}},\tilde S'_{\nu\bm k\ulin{\bm k}}]=i\sum_\rho \varepsilon_{\mu\nu\rho}\tilde S'_{\rho\bm k\underline{\bm k}}~.\label{eq.commuSStd}
\end{eqnarray}
\par
The spin-1 triplet states for $S'_{z\bm k\ulin{\bm k}}$ can be written as
\begin{eqnarray}
&&\ket{1'}_{\bm k\underline{\bm k}}=\ket{\upa'\upa'}_{\bm k\underline{\bm k}},\\
&&\ket{0'}_{\bm k\underline{\bm k}}=\frac{1}{\sqrt{2}}(\ket{\upa'\dna'}_{\bm k\underline{\bm k}}-\ket{\dna'\upa'}_{\bm k\underline{\bm k}}),\\
&&\ket{-1'}_{\bm k\underline{\bm k}}=\ket{\dna'\dna'}_{\bm k\underline{\bm k}}~.
\end{eqnarray}
They satisfy $S'_{z\bm k\underline{\bm k}}\ket{m'}_{\bm k\underline{\bm k}} =m\ket{m'}_{\bm k\underline{\bm k}}$ ($m=1,0,-1$). They have even parity under $\mathcal C_{\bm k\ulin{\bm k}}$: $\mathcal C_{\bm k\ulin{\bm k}}\ket{m'}_{\bm k\underline{\bm k}}=\ket{m'}_{\bm k\underline{\bm k}}$. 
\par
The spin-0 singlet state $\ket{\tilde 0'}_{\bm k\ulin{\bm k}}$ is given as
\begin{eqnarray}
\ket{{\tilde 0}'}_{\bm k\ulin{\bm k}}=\frac{1}{\sqrt{2}}(\ket{\upa'\dna'}_{\bm k\ulin{\bm k}}+\ket{\dna'\upa'}_{\bm k\ulin{\bm k}})~.
\end{eqnarray}
It satisfies $S'_{\mu\bm k\ulin{\bm k}}\ket{\tilde 0'}_{\bm k\ulin{\bm k}}=0$ ($\mu=x,y,z$). It has odd parity under $\mathcal C_{\bm k\ulin{\bm k}}$: $\mathcal C_{\bm k\ulin{\bm k}}\ket{\tilde 0'}_{\bm k\ulin{\bm k}}=-\ket{\tilde 0'}_{\bm k\ulin{\bm k}}$. 
\par
$S'^\pm_{\bm k\ulin{\bm k}}=S'_{x\bm k\ulin{\bm k}}\pm iS'_{y\bm k\ulin{\bm k}}$ are the ladder operators for the spin-1 triplet states $\{\ket{m'}_{\bm k\ulin{\bm k}}\}$:
\begin{eqnarray}
&&S'^-_{\bm k\ulin{\bm k}}\ket{1'}_{\bm k\ulin{\bm k}}=S'^+_{\bm k\ulin{\bm k}}\ket{-1'}_{\bm k\ulin{\bm k}}=\sqrt{2}\ket{0'}_{\bm k\ulin{\bm k}}~,\label{eq.ladderopd1}\\
&&S'^+_{\bm k\ulin{\bm k}}\ket{1'}_{\bm k\ulin{\bm k}}=S'^-_{\bm k\ulin{\bm k}}\ket{-1'}_{\bm k\ulin{\bm k}}=0~,\label{eq.ladderopd2}\\
&&S'^\pm_{\bm k\ulin{\bm k}}\ket{0'}_{\bm k\ulin{\bm k}}=
\left\{
\begin{array}{l}
\sqrt{2}\ket{1'}_{\bm k\ulin{\bm k}}~,\\
\sqrt{2}\ket{-1'}_{\bm k\ulin{\bm k}}~.
\end{array}
\right.\label{eq.ladderopd3}
\end{eqnarray}
$\tilde{S}'^\pm_{\bm k\ulin{\bm k}}=\tilde{S}'_{x\bm k\ulin{\bm k}}\pm i\tilde{S}'_{y\bm k\ulin{\bm k}}$ transform $\ket{m'}_{\bm k\ulin{\bm k}}$ into $\ket{\tilde 0'}_{\bm k\ulin{\bm k}}$ and {\it vice versa}: 
\begin{eqnarray}
&&\tilde{S}'^-_{\bm k\ulin{\bm k}}\ket{1'}_{\bm k\ulin{\bm k}}=-\tilde{S}'^+_{\bm k\ulin{\bm k}}\ket{-1'}_{\bm k\ulin{\bm k}}=-\sqrt{2}\ket{\tilde{0}'}_{\bm k\ulin{\bm k}}~,\label{eq.Stild1}\\
&&\tilde{S}'^+_{\bm k\ulin{\bm k}}\ket{1'}_{\bm k\ulin{\bm k}}=\tilde{S}'^-_{\bm k\ulin{\bm k}}\ket{-1'}_{\bm k\ulin{\bm k}}=\tilde{S}'^\pm_{\bm k\ulin{\bm k}}\ket{0'}_{\bm k\ulin{\bm k}}=0~,\label{eq.Stild2}\\
&&\tilde{S}'^+_{\bm k\ulin{\bm k}}\ket{\tilde{0}'}_{\bm k\ulin{\bm k}}=-\sqrt{2}\ket{1'}_{\bm k\ulin{\bm k}},\\
&&\tilde{S}'^-_{\bm k\ulin{\bm k}}\ket{\tilde{0}'}_{\bm k\ulin{\bm k}}=\sqrt{2}\ket{-1'}_{\bm k\ulin{\bm k}}~.\label{eq.Stild3}
\end{eqnarray}
\par
Given the MF Hamiltonian in Eq.~(11), using Eqs.~(\ref{eq.commuSSd}) and (\ref{eq.commuSStd}), one obtains
\begin{eqnarray}
\left[\mathcal H_{\rm MF},S'^{\pm}_{\bm k\ulin{\bm k}}\right]=\pm 2E_{\bm k}S'^{\pm}_{\bm k\ulin{\bm k}}~,\label{eq.commuHMF1}\\
\left[\mathcal H_{\rm MF},{\tilde S}'^{\pm}_{\bm k\ulin{\bm k}}\right]=\pm 2E_{\bm k}\tilde{S}'^{\pm}_{\bm k\ulin{\bm k}}~,\label{eq.commuHMF2}\\
\left[\mathcal H_{\rm MF},S'^{\pm}_{\bm k_F}\right]=\pm 2\Delta_0 S'^{\pm}_{\bm k_F}~.\label{eq.commuHMF3}
\end{eqnarray}
$S'^{\pm}_{\bm k\ulin{\bm k}}$, ${\tilde S}'^{\pm}_{\bm k\ulin{\bm k}}$, and $S'^{\pm}_{\bm k_F}$ are the raising and lowering operators for the eigenstates of $\mathcal H_{\rm MF}$. $\ket{e_{\bm k\ulin{\bm k}}^0}=S_{\bm k\ulin{\bm k}}'^+\ket{\Psi}/\sqrt{2}$ and $\ket{e_{\bm k\ulin{\bm k}}^{\tilde 0}}={\tilde S}_{\bm k\ulin{\bm k}}'^+\ket{\Psi}/\sqrt{2}$ are the degenerate excited states with excitation energy $2E_{\bm k}$ that involves a single pair of bogolons. They have even and odd parity under $\mathcal C_{\bm k\ulin{\bm k}}$, respectively. $\ket{e_{\bm k_F}}={S'}^+_{\bm k_F}\ket{\Psi}$ is an excited state with excitation energy $2\Delta_0$ that has odd parity under $\mathcal C_{\bm k_F}$.

\section{Holstein-Primakoff Theory applied to $\mathcal H_I$}

In this Section, we apply the Holstein-Primakoff theory to $\mathcal H_I$ in order to derive the Higgs mode.
Substituting Eq.~(\ref{eq.Sprime}) into $\mathcal H_I$, it can be written as
\begin{eqnarray}
&&\mathcal H_I=\sideset{}{^{'}}\sum_{\bm k}2\xi_{\bm k}(-\cos\varphi_{\bm k}S'_{z\bm k\ulin{\bm k}}+\sin\varphi_{\bm k}S'_{x\bm k\ulin{\bm k}})\nonumber\\
&&-g\left\{\sideset{}{^{'}}\sum_{\bm k,\bm k'}(\cos\varphi_{\bm k}\cos\varphi_{\bm k'}S'_{x\bm k\ulin{\bm k}}S'_{x\bm k'\ulin{\bm k}'}
+\sin\varphi_{\bm k}\cos\varphi_{\bm k'}\right.\nonumber\\
&&\times\{S'_{z\bm k\ulin{\bm k}},S'_{x\bm k'\ulin{\bm k}'}\}+\sin\varphi_{\bm k}\sin\varphi_{\bm k'}S'_{z\bm k\ulin{\bm k}}S'_{z\bm k'\ulin{\bm k}'})\nonumber\\
&&+2\sideset{}{^{'}}\sum_{\bm k}\sum_{\bm k_F}(-\sin\varphi_{\bm k}S'_{z\bm k\ulin{\bm k}}-\cos\varphi_{\bm k}S'_{x\bm k\ulin{\bm k}})S_{x\bm k_F}\nonumber\\
&&\left.+\sum_{\bm k_F,\bm k_F'}S_{x\bm k_F}S_{x\bm k_F'}\right\}. \label{eq.HBCSdash}
\end{eqnarray}
Since the eigenstates of $\mathcal H_I$ can be characterized by the parity under $\mathcal C_{\bm k_F}$, we can treat $S_{x\bm k_F}$ as a $c$-number $S_{x\bm k_F}=\pm1/2$.
We quantize spin fluctuations around the ground state $\ket{\Psi}$ by the Holstein-Primakoff transformation [45]:
\begin{eqnarray}
&&S'^+_{\bm k\ulin{\bm k}}=\sqrt{2S}\gamma_{\bm k}^\dagger \sqrt{1-\frac{\gamma_{\bm k}^\dagger\gamma_{\bm k}}{2S}}~,\label{eq.HPtrans1}\quad
S'^-_{\bm k\ulin{\bm k}}=(S'^+_{\bm k\ulin{\bm k}})^\dagger~,\label{eq.HPtrans2}\\
&&S'_{z\bm k\ulin{\bm k}}=-\left(S-\gamma_{\bm k}^\dagger\gamma_{\bm k}\right)~,\label{eq.HPtrans3}
\end{eqnarray}
where $\gamma_{\bm k}^\dagger$ and $\gamma_{\bm k}$ denote, respectively, the creation and annihilation operators of a boson that represents spin fluctuations. They satisfy the usual commutation relations  $[\gamma_{\bm k},\gamma_{\bm k'}^\dagger]=\delta_{\bm k,\bm k'}$ and $[\gamma_{\bm k},\gamma_{\bm k'}]=[\gamma_{\bm k}^\dagger,\gamma_{\bm k'}^\dagger]=0$. 
In view of the fact that the creation operator of the Higgs mode derived from $\mathcal H_{\rm BCS}$ in Ref.~[42] has even parity under $\mathcal C_{\bm k\ulin{\bm k}}$, we restrict the Hilbert space spanned by even-parity states under $\mathcal C_{\bm k\ulin{\bm k}}$, in which $\bm S'_{\bm k\ulin{\bm k}}$ represents a spin-1 operator. We thus set $S=1$.
If spin fluctuation is small ($\gamma_{\bm k}^\dagger\gamma_{\bm k}\ll 1$), $S_{\bm k\ulin{\bm k}}'^\pm$ can be approximated as $S_{\bm k\ulin{\bm k}}'^+\simeq \sqrt{2}\gamma_{\bm k}^\dagger$ and $S_{\bm k\ulin{\bm k}}'^-\simeq \sqrt{2}\gamma_{\bm k}$.
\par
We expand Eq.~(\ref{eq.HBCSdash}) in terms of $\gamma_{\bm k}$ and $\gamma_{\bm k}^\dagger$. The zeroth and first-order terms read
\begin{eqnarray}
\mathcal H_I^{(0)}&=&-2\sideset{}{^{'}}\sum_{\bm k}\frac{\xi_{\bm k}^2}{E_{\bm k}}-\frac{\Delta_0^2}{g}~,\\
\mathcal H_I^{(1)}&=&\sqrt{2}\sideset{}{^{'}}\sum_{\bm k}(\xi_{\bm k}\sin\varphi_{\bm k}+\Delta_0\cos\varphi_{\bm k})(\gamma_{\bm k}+\gamma_{\bm k}^\dagger)~.
\end{eqnarray}
Here, we have used the approximation
\begin{equation}
\Delta_0=g(\sideset{}{^{'}}\sum_{\bm k}\sin\varphi_{\bm k}+\sum_{\bm k_F}S_{x\bm k_F})\simeq g\sideset{}{^{'}}\sum_{\bm k}\sin\varphi_{\bm k}~.\label{eq.gapap}
\end{equation}
Since $\mathcal H_I^{(1)}$ should vanish, we obtain $\sin\varphi_{\bm k}=\Delta_0/E_{\bm k}$ and $\cos\varphi_{\bm k}=-\xi_{\bm k}/E_{\bm k}$. Equation~(\ref{eq.gapap}) thus reduces to the gap equation  
\begin{equation} 
1=g\sideset{}{^{'}}\sum_{\bm k}\frac{1}{E_{\bm k}}~.\label{eq.gapeq}
\end{equation}
\par
The second-order term reads
\begin{eqnarray}
\mathcal H_I^{(2)}&=&2\sideset{}{^{'}}\sum_{\bm k}E_{\bm k}\gamma_{\bm k}^\dagger\gamma_{\bm k}-\frac{g}{2}\sideset{}{^{'}}\sum_{\bm k,\bm k'}\cos\varphi_{\bm k}\cos\varphi_{\bm k'}\nonumber\\
&&\times(\gamma_{\bm k}\gamma_{\bm k'}+\gamma_{\bm k}\gamma_{\bm k'}^\dagger+\gamma_{\bm k}^\dagger\gamma_{\bm k'}+\gamma_{\bm k}^\dagger\gamma_{\bm k'}^\dagger)~.\label{eq.H2}
\end{eqnarray}
We diagonalize $\mathcal H_I^{(2)}$ by the Bogoliubov transformation
\begin{eqnarray}
\beta_{\lambda}=\sum_{\bm k}(U_{\lambda\bm k}^*\alpha_{\bm k}+V_{\lambda\bm k}^*\alpha_{\bm k}^\dagger)~,\label{eq.bosonbogoliubov1}\\
\beta_{\lambda}^\dagger=\sum_{\bm k}(U_{\lambda\bm k}\alpha_{\bm k}^\dagger+V_{\lambda\bm k}\alpha_{\bm k})~,\label{eq.bosonbogoliubov2}
\end{eqnarray}
where $\lambda$ labels the normal modes. The bosonic operator $\beta_\lambda$ satisfies the commutation relations
\begin{eqnarray}
&&[\beta_\lambda,\beta_{\lambda'}^\dagger]=\sum_{\bm k}(U_{\lambda\bm k}^*U_{\lambda'\bm k}-V_{\lambda\bm k}^*V_{\lambda'\bm k})=\delta_{\lambda,\lambda'}~,\label{eq.normalcondition}\\
&&[\beta_\lambda^\dagger,\beta_{\lambda'}^\dagger]=\sum_{\bm k}(-U_{\lambda\bm k}V_{\lambda'\bm k}+V_{\lambda\bm k}U_{\lambda'\bm k})=0~.
\end{eqnarray}
One can derive the inverse transformation from Eqs.~(\ref{eq.bosonbogoliubov1}) and (\ref{eq.bosonbogoliubov2}) as
\begin{eqnarray}
\gamma_{\bm k}=\sum_{\lambda}(U_{\lambda\bm k}\beta_\lambda-V_{\lambda\bm k}^*\beta_{\lambda}^\dagger)~,\\
\gamma_{\bm k}^\dagger=\sum_{\lambda}(U_{\lambda\bm k}^*\beta_\lambda^\dagger-V_{\lambda\bm k}\beta_{\lambda})~.
\end{eqnarray}
\par
Assuming that the second order term is diagonalized as $\mathcal H_I^{(2)}=\sum_\lambda\omega_\lambda\beta_\lambda^\dagger\beta_\lambda+{\rm const.}$, we obtain
\begin{eqnarray}
[\gamma_{\bm k},\mathcal H_I^{(2)}]=\sum_{\lambda}\omega_{\lambda}(U_{\lambda\bm k}\beta_{\lambda}+V_{\lambda\bm k}^*\beta_{\lambda}^\dagger)~.
\label{eq.commute1}
\end{eqnarray}
On the other hand, using Eq.~(\ref{eq.H2}), we obtain
\begin{eqnarray}
&&[\gamma_{\bm k},\mathcal H_I^{(2)}]\nonumber\\
&&=\sum_{\lambda}\left\{\left(2E_{\bm k}U_{\lambda\bm k}-g\sideset{}{^{'}}\sum_{\bm k'}\cos\varphi_{\bm k}\cos\varphi_{\bm k'}(U_{\lambda\bm k'}-V_{\lambda\bm k'})\right)\beta_{\lambda}\right.\nonumber\\
&&\left.+\left(-2E_{\bm k}V^*_{\lambda\bm k}-g\sideset{}{^{'}}\sum_{\bm k'}\cos\varphi_{\bm k}\cos\varphi_{\bm k'}(U^*_{\lambda\bm k'}-V^*_{\lambda\bm k'})\right)\right\}\beta^\dagger_{\lambda}~.
\label{eq.commute2}
\end{eqnarray}
For Eqs.~(\ref{eq.commute1}) and (\ref{eq.commute2}) to be consistent, $X_{\lambda\bm k}$ and $Y_{\lambda\bm k}$ should satisfy
\begin{eqnarray}
&&2E_{\bm k}U_{\lambda\bm k}-g(e_\lambda-f_\lambda)\cos\varphi_{\bm k}=\omega_{\lambda}U_{\lambda\bm k}~,\label{eq.equationX}\\
&&-2E_{\bm k}V_{\lambda\bm k}-g(e_\lambda-f_\lambda)\cos\varphi_{\bm k}=\omega_{\lambda}V_{\lambda\bm k}~,\label{eq.equationY}
\end{eqnarray}
where the coefficients $e_\lambda$ and $f_\lambda$ are given by
\begin{eqnarray}
e_\lambda=\sideset{}{^{'}}\sum_{\bm k}\cos\varphi_{\bm k}U_{\lambda \bm k}~,\label{eq.e}\\
f_\lambda=\sideset{}{^{'}}\sum_{\bm k}\cos\varphi_{\bm k}V_{\lambda \bm k}~.\label{eq.f}
\end{eqnarray}
If $e_\lambda-f_\lambda\neq 0$, Eqs~(\ref{eq.equationX}) and (\ref{eq.equationY}) can be formally solved as
\begin{eqnarray}
U_{\lambda\bm k}=g\frac{(e_\lambda-f_\lambda)\cos\varphi_{\bm k}}{2E_{\bm k}-\omega_\lambda}~,\label{eq.formalsol1}\\
V_{\lambda\bm k}=-g\frac{(e_\lambda-f_\lambda)\cos\varphi_{\bm k}}{2E_{\bm k}+\omega_\lambda}~.\label{eq.formalsol2}
\end{eqnarray}
We omit $\lambda$ below.
\par
Substituting Eqs.~(\ref{eq.formalsol1}) and (\ref{eq.formalsol2}) into Eqs.~(\ref{eq.e}) and (\ref{eq.f}), we obtain
\begin{eqnarray}
1-4g\sideset{}{^{'}}\sum_{\bm k}\frac{\xi_{\bm k}^2}{E_{\bm k}^2}\frac{E_{\bm k}}{4E_{\bm k}^2-\omega^2}=0~.
\end{eqnarray}
The above equation has a solution $\omega=2\Delta_0$, for which it reduces to the MF gap equation (\ref{eq.gapeq}). 
We thus obtain 
\begin{eqnarray}
U_{\bm k}&=&\frac{{B}\cos\varphi_{\bm k}}{2\Delta_0-2E_{\bm k}},\quad V_{\bm k}=\frac{{B}\cos\varphi_{\bm k}}{2\Delta_0+2E_{\bm k}}~,\label{eq.Higgsamplitude}
\end{eqnarray}
where $B$ is the normalization constant. $B$ is determined by the normalization condition (\ref{eq.normalcondition}) as
\begin{eqnarray}
B=\frac{1}{\sqrt{\sum_{\bm k}'\frac{\Delta_0}{E_{\bm k}\xi_{\bm k}^2}}}~.
\end{eqnarray}
The creation operator for the collective excitation with $\omega=2\Delta_0$ is thus given by
\begin{eqnarray}
\beta_{\rm H}^\dagger=B\sideset{}{^{'}}\sum_{\bm k}\frac{\xi_{\bm k}}{E_{\bm k}}\left(\frac{\gamma_{\bm k}^\dagger}{2\Delta_0-2E_{\bm k}}+\frac{\gamma_{\bm k}}{2\Delta_0+2E_{\bm k}}\right)~.
\label{eq.Higgscreation1}
\end{eqnarray}
$\beta_{\rm H}^\dagger$ coincides with the creation operator of the Higgs mode derived from $\mathcal H_{\rm BCS}$ (See Eq.~(C31) in Ref.~[42]) using $\alpha_{\bm k}^\dagger-\alpha_{\ulin{\bm k}}^\dagger=\sqrt{2}\gamma_{\bm k}^\dagger$. Here, $\alpha_{\bm k}^\dagger$ is a creation operator of a boson that describes fluctuations of the spin-1/2 operator ${\bm S}'_{\bm k}$ (See Eqs.~(C2) and (C3) in Ref.~[42]). 
\par
In the limit of small fluctuations ($\gamma_{\bm k}^\dagger\gamma_{\bm k}\ll 1$), using $\gamma_{\bm k}^\dagger\simeq S'^+_{\bm k\ulin{\bm k}}/\sqrt{2}$ and $\gamma_{\bm k}\simeq S'^-_{\bm k\ulin{\bm k}}/\sqrt{2}$, the creation operator for the Higgs mode can be written as 
\begin{eqnarray}
\beta_{\rm H}^\dagger=\frac{B}{\sqrt{2}}\sideset{}{^{'}}\sum_{\bm k}\frac{\xi_{\bm k}}{E_{\bm k}}\left(\frac{S'^+_{\bm k\ulin{\bm k}}}{2\Delta_0-2E_{\bm k}}+\frac{S'^-_{\bm k\ulin{\bm k}}}{2\Delta_0+2E_{\bm k}}\right)~.
\label{eq.Higgscreation2}
\end{eqnarray}
We omit the normalization constant in Eq.~(13) in the main text.
The excited state with a single Higgs mode can be written in terms of $\ket{e^0_{\bm k\ulin{\bm k}}}$ as
\begin{equation}
\beta_{\rm H}^\dagger\ket{\Psi}=B\sideset{}{^{'}}\sum_{\bm k}\frac{\xi_{\bm k}}{E_{\bm k}}\frac{1}{2\Delta_0-2E_{\bm k}}\ket{e^0_{\bm k\ulin{\bm k}}}~.
\end{equation}

\end{document}